\documentclass[a4paper,11pt]{article}
\pdfoutput=1 % if your are submitting a pdflatex (i.e. if you have
\usepackage{jcappub} % for details on the use of the package, please
\usepackage[T1]{fontenc} % if needed
\usepackage{cleveref}
\usepackage[svgnames]{xcolor}
\usepackage{enumitem}
\usepackage[thinlines]{easytable}
\usepackage{caption}

\newcommand{\p}{\partial}

\title{\boldmath Are $f(R, {\rm Matter})$ theories really relevant to cosmology? }

\author[a,b]{Osmin Lacombe,}
\author[c,d]{Shinji Mukohyama,}
\author[c,e]{and Josef Seitz}

\affiliation[a]{Dipartimento di Fisica e Astronomia, Universit\`a di Bologna,
via Irnerio 46, 40126, Bologna, Italy}
\affiliation[b]{INFN, Sezione di Bologna, viale Berti Pichat 6/2, 40127 Bologna, Italy}
\affiliation[c]{Center for Gravitational Physics and Quantum Information, Yukawa Institute for Theoretical Physics, Kyoto University, Sakyo-ku, Kyoto 606-8502, Japan}
\affiliation[d]{Kavli Institute for the Physics and Mathematics of the Universe, the University of Tokyo Institutes for Advanced Study, the University of Tokyo, Kashiwa, Chiba, 277-8583, Japan}
\affiliation[e]{Arnold Sommerfeld Center, Ludwig-Maximilians University, Theresienstraße 37, 80333,\\ Munich, Germany}

\emailAdd{deriusosmin.lacombe@unibo.it}
\emailAdd{shinji.mukohyama@yukawa.kyoto-u.ac.jp}
\emailAdd{josef.seitz@physik.uni-muenchen.de}

\abstract{We examine $f(R, \text{Matter})$ theories that directly couple the curvature $R$ or $R_{\mu\nu}$ with the matter sector in the action, in addition to the universal coupling. We argue that if the matter sector includes the Standard Model (SM), such theories are either inconsistent or already excluded by experiments unless they are a rewriting of $f(R)$ gravity or general relativity. If these theories genuinely couple the SM to curvature, they suffer from the presence of ghost states at energies within their domain of application for cosmological purposes. Therefore, we raise questions about their relevance to cosmology. Moreover, if such theories do not include the SM, they should just be seen as scalar-tensor, vector-tensor, \ldots, theories, depending on the additional degrees of freedom. They should thus be studied accordingly.}

\makeatletter
\gdef\@fpheader{}
\makeatother

\begin{document}

\begin{flushright} {YITP-23-149\\IPMU23-0044}  \end{flushright}
\vspace{-1.4cm}

\maketitle
\flushbottom

~\\

\section{Introduction}
\label{sec:intro}

General Relativity (GR) is the simplest possible covariant gravitational theory with two local physical degrees of freedom. Its dynamics are governed by the action
\begin{equation} \label{EH}
    S = \int d^4x \sqrt{-g} \left(\frac{1}{16\pi G} R + \mathcal{L}_m\right),
\end{equation}
with $\mathcal{L}_m$ the Lagrangian of the matter fields. 

While GR has proven to be experimentally very successful, there have been numerous attempts to modify or extend it since its discovery. These endeavours are justified by the fact that, when coupled to the Standard Model (SM) of particle physics, GR falls short in  explaining certain phenomena such as the existence of dark energy, dark matter, or inflation. Addressing these phenomena thus requires either going beyond the SM of particle physics while retaining GR, solely modifying the laws of gravity, or doing both. One of the simplest modifications of gravity is to make use of an arbitrary function $f(R)$ of the Ricci scalar $R$. Such modification can be motivated by quantum corrections and provides an inflationary model for a certain choice of $f$ \cite{Starobinsky:1980te,Vilenkin:1985md}, provided that other corrections, such as the Weyl squared term, are suppressed. 

Explaining observations is the primary motivation behind modified gravity theories; however, these theories must be theoretically consistent and tractable. For instance, they should be ghost-free within their regime of validity and weakly coupled, as predictions become difficult in strongly coupled theories. Modifying the predictions of GR in a specific regime while simultaneously maintaining consistency in that same regime is a challenging but achievable task. To name a few, examples of such possibilities include $f(R)$ gravity  \cite{Starobinsky:1980te,Strominger:1984dn,Woodard:2006nt}, scalar-tensor theories \cite{Damour:1992we} or extra dimensions models \cite{Randall:1999vf,Dvali:2000hr}.

Over the past decade, a new class of modified gravity has become popular. It uses non-universal couplings of gravity with the matter sector, via the curvature tensor. We henceforth call this class $f(R,\rm{Matter})$ theories, and gather under this name $f(R,T)$ gravity \cite{Harko:2011kv} and its variants such as $f(R, \mathcal{L}_m)$\cite{Harko:2010mv}, $f(R, T, T_{\mu \nu} R^{\mu \nu})$ \cite{Odintsov:2013iba,Haghani:2013oma}, $f(R,T_{\mu \nu} T^{\mu \nu})$\cite{Katirci:2013okf} and $f(R, G_{\mu \nu}T^{\mu \nu})$ \cite{Marciu:2023jvs}. 
To be fully explicit, the theories of this class are defined by replacing the Ricci scalar $R$ of the Einstein-Hilbert action \eqref{EH} by a function of the curvature tensor and the matter Lagrangian:
\begin{equation}
    S = \int d^4 x\sqrt{-g} \left(\frac{1}{16\pi G} f(R,\rm{Matter})+\mathcal{L}_m\right),
\end{equation}
 A specific model of the $f(R,\rm{Matter})$ class thus corresponds to choices of the function $f$ and its arguments, namely the function of the curvature -- usually $R$ or $R_{\mu\nu}$, and of the matter sector -- usually the matter Lagrangian $\mathcal{L}_m$, the stress-energy tensor $T_{\mu\nu}$ or its trace $T$. The explicit action would thus have the general form shown in \eqref{GeneralAction}. While many choices of $f$ have been studied in the past, the focus has been mostly on observational effects. Such studies usually first assume that the matter energy-momentum tensor $T^{\mu \nu}$ takes the perfect fluid form, and they subsequently study the field equations in a FLRW, nearly de Sitter, or flat Universe. This approach has been used to claim that $f(R,\rm{Matter})$ gravity can explain various cosmological and astrophysical observations. In the context of cosmology, those applications range from inflation \cite{Oikonomou:2023bmn,Gamonal:2020itt,Chen:2022dyq,Mohammadi:2023sqy,Taghavi:2023ptn,Bhattacharjee:2020jsf} to dark matter \cite{Zaregonbadi:2016xna,Mohan:2022kvb, Shabani:2022buw} and dark energy \cite{Sun:2015yga, Myrzakulov:2012ug, Maurya:2022pzw}. 

While previous studies have explored in great detail the consequences of $f(R,\rm{Matter})$ theories in the perfect fluid approximation,  little attention has been given to the question of the theoretical consistency of $f(R,\rm{Matter})$ theories. See however \cite{Ayuso:2014jda} for some initial comments on this point. Given the difficulty of consistently modifying GR, this question is well motivated. As the $f(R,\rm{Matter})$ class is quite broad, the answer could, {\it a priori}, differ depending on the model. However, we believe that all of them suffer from some serious issues, raising questions about their relevance to cosmology.

In this paper, we argue that $f(R,\rm{Matter})$ theories fall in one of the following four categories:  
\begin{enumerate}[label=\alph*)]
    \item they are equivalent to $f(R)$ + matter Lagrangian,
    \item they only modify the matter sector with respect to GR or $f(R)$,
    \item they are just a rewriting of scalar-tensor, vector-tensor, etc \ldots, theories 
    \item they suffer from the presence of ghosts at energies within their domain of application to cosmology.
\end{enumerate}
We point out that the categories a), b), and c) are not problematic {\it per se} but do not correspond to a genuine coupling of the matter sector with the curvature tensor. For instance, models falling in category b) should be thought of as Beyond the Standard Model models, instead of modifications of gravity. As such, they must not lead to undetected effects at energies below the weak scale. The main focus of this paper is thus to explain that all genuine $f(R,\rm{Matter})$ theories actually fall in the category d).

The outline of the paper is as follows: in \Cref{sec:generalarguments}, we define the necessary conditions for $f(R,\rm{Matter})$ theories to be new consistent modifications of gravity.  We illustrate these conditions with particular forms of $f(R,\rm{Matter})$, supporting the earlier classification. This section alone should convince the reader of the challenges in constructing cosmologically relevant theories of this kind. Moving to \Cref{sec:specifics}, we delve into a particular case: $f(R,T)$ gravity in a low curvature expansion. We first explicitly present the particle physics constraints when $f(R,T) = R+f_0(T)$, a pure matter sector modification. Subsequently, we study $R h(T)$ matter-curvature couplings and reveal the presence of  a ghost at scales in the regime of interest for cosmological applications. The paper concludes in \Cref{sec:conclusions} with a short discussion on the perfect fluid assumption in $f(R,\rm{Matter})$ theories before summarizing our results.

\section{General Arguments}
\label{sec:generalarguments}
In this section, we start by listing the conditions modified gravity theories must fulfill  to be considered as a new modified theory of gravity in the $f(R,\rm Matter)$ class. These conditions are grouped in \Cref{subsec:requirements}. We then show in \Cref{subsec:combinationarguments} how combining these arguments rules out the possibility of constructing consistent new models through an action of the $f(R,{\rm Matter})$ form. To be fully explicit,  we study actions of the form:
\begin{align}
S&=\frac{M_P^2}2\int  \, d^4x \sqrt{g} \, \big\{ f(R) + f_1(R,\mathcal{L}_m) +f_2(R,T) + f_3(R_{\mu\nu}T^{\mu\nu})  \big\} + \int d^4x \sqrt{-g} \, \mathcal{L}_m, \label{GeneralAction}
\end{align}
where $M_P^2=1/(8\pi G)$ is the (reduced) squared Planck mass expressed in terms of Newton's constant $G$ and
 $T$ denotes the trace of the energy-momentum tensor, defined with respect to the matter Lagrangian $\mathcal{L}_m$:
\begin{equation}
    T_{\mu \nu} = -\frac{2}{\sqrt{-g}}\frac{\delta \sqrt{-g} \mathcal{L}_m}{\delta g^{\mu \nu}}.
\end{equation}
The arguments we give below do not apply to the cases of general relativity (GR) or $f(R)$ theories, which correspond to $f_1=f_2=f_3=0$.

\subsection{Conditions for valid theories}
\label{subsec:requirements}

We consider the following list of requirements to evaluate the relevance of theories modelled by a Lagrangian of the above form.
\begin{enumerate}

\item {\bf The matter sector containing the Standard Model}
\label{condition:notscalartensor}
The first point we address here is related to the definition of the matter Lagrangian $\mathcal{L}_m$ or stress-energy tensor $T^{\mu\nu}$ used in the action \eqref{GeneralAction}. The functions $\mathcal{L}_m$, $T$ or yet $T^{\mu\nu}$ appearing in $f_2$, $f_3$ and $f_4$ must refer to the true matter sector. By this, we mean that they should include (at least) the Standard Model (SM) fields, such as the photon, Higgs field \ldots 

This requirement rules out theories where the Lagrangian $\mathcal{L}_m$ or the trace tensor $T$ refer to fields, $e.g.$ scalar fields,  which are not directly related to our visible matter sector. Indeed, in that case, the action \cref{GeneralAction} can be seen simply as a scalar-tensor theory of gravity \cite{Damour:1992we}, and has nothing to do with a coupling of matter to the gravity sector.

\item {\bf Not simply a modification of the matter sector}
\label{condition:matterismatter}  This second requirement is related to the previous one, but applies to somewhat different cases.  We require that the theory under consideration does not only constitute a modification of the matter sector.  

 This rules out all the theories where only the stress-energy tensor and/or the matter Lagrangian appear in the action \eqref{GeneralAction}, namely those with $f_3=0$ and possibly $f_1(\mathcal{L}_m,R)=f_1(\mathcal{L}_m)$, $f_2(R,T)=f_2(T)$. Indeed, in such cases, the action reduces to a $f(R)$ theory of gravity coupled universally to a better identified (true) matter Lagrangian of the form
 \begin{equation}
\mathcal{L}_m^{(true)}=\mathcal{L}_m+f_1(\mathcal{L}_m)+f_2(T). \label{LmTrue}
 \end{equation}
Hence these theories can be seen as theories that only define a new matter Lagrangian, namely theories of exotic matter with Lagrangian $\mathcal{L}_m^{(true)}$\cite{Fisher:2019ekh}. This kind of theory will in general be in disagreement with particle accelerator experiments if the original $\mathcal{L}_m$ includes the SM Lagrangian. Moreover the form of \eqref{LmTrue} highly constrains operators of different mass dimensions, relating their couplings in a very non-trivial way.

In any case, such theories cannot be seen as modifications of the gravity sector.

\item{\bf Relevant energy scales and parameters} \label{condition:energyscale} We demand that the theory we are studying is valid until a certain ultraviolet (UV) scale $\Lambda$. Hence, even if the Lagrangian under study is seen as low energy effective Lagrangian, the theory it describes should be consistent until that UV scale. It is thus in this very domain of validity that the theory should be relevant for phenomenology. 

Theories modelled by a Lagrangian of the form of \eqref{GeneralAction} are in general self-consistent only until a certain UV scale $ \Lambda(\lambda_1,\lambda_2,\lambda_3,\ldots)$, as is mentioned in the next requirement and explained in more details in the rest of the paper. This UV scale depends on the parameters $\lambda_i, \ldots$ appearing in the functions $f_j$, which parametrize couplings between matter and gravity. These are the exact parameters that determine how much the theory differs from GR or $f(R)$ theories, at the level of the Lagrangian and above all in observable quantities.  

Hence, we ask that the theories under study have observable deviations from GR or $f(R)$ theories in their own domains of validity. Put another way, the $\lambda_i$ parameters must be such that the $f_j$ couplings have observable effects below a certain UV scale $\Lambda(\{\lambda_i\})$ where the theory remains self-consistent.

\item{\bf Absence of ghosts}
\label{condition:2ndorder-eom}
This requirement makes more precise what we mean by the self-consistency or validity of a theory. The main concern for theories of the form of \cref{GeneralAction} are instabilities coming from ghosts, fields with wrong kinetic sign. Such instabilities render the energy unbounded from below and violate the unitarity of the theory. A famous example is the Ostrogradsky ghost \cite{Ostrogradsky:1850fid} in the presence of higher-derivative operators. 

Ghost degrees of freedom arise when higher-order derivative terms are present in the Lagrangian. Such terms (almost) always lead to equations of motion with higher-than-two time derivatives. From the classical point of view, solving these equations requires more than two initial conditions, hence signaling the presence of these bad additional degrees of freedom. 

Modifications of gravity with higher-derivative terms in the Lagrangian have been studied extensively. There are two ways to avoid the presence of ghosts. The first is to consider specific forms of higher-derivative scalar-tensor Lagrangians, keeping second-order equations of motion. The specific forms of these Lagrangian including the metric $g_{\mu\nu}$ as well as a scalar field $\phi$ have been classified in the famous work of Horndeski  \cite{Horndeski:1974wa}. The second is to consider a Lagrangian which albeit leading to higher-order-derivative equations of motion, does not contain any ghosts. This is the case when the terms of the Lagrangian satisfy certain degeneracy conditions. Diverse such theories have been developed gradually until explicit degeneracy conditions have been found, grouping them all together under the name of DHOST theories \cite{Langlois:2015cwa}. 

By requiring the absence of ghosts in the effective theories described by an action of the form of \cref{GeneralAction}, we thus ask them to satisfy either the Horndeski form or the degeneracy conditions of DHOST theories. More precisely, this requirement must hold up to higher-derivative terms that break the degeneracy condition at and above a certain UV scale -- the so called scordatura terms \cite{Motohashi:2019ymr,Gorji:2020bfl,Gorji:2021isn,DeFelice:2022xvq}.

\item {\bf Compatibility with observations}
\label{condition:obsconstraints}
Last but not least, we require that the theory under consideration does not lead to observable quantities in contradiction with experiments. This refers in particular to particle physics experiments, feeding our knowledge about the form of the matter Lagrangian below certain energies (the SM Lagrangian), cosmological observations, \ldots

\end{enumerate}

We argue that the above natural conditions are sufficient to show that theories of action \eqref{GeneralAction} -- except for GR and $f(R)$ theories -- are not good candidates for applications in cosmology. Several cases can appear. In the first cases, the matter Lagrangian under consideration does not describe our proper matter, so that the theory is just a rewriting of a scalar-tensor theory or a modification of the matter sector. In other cases, the theory is not self-consistent, due to the presence of ghost instabilities, below the UV scale of interest for its phenomenological application.

\subsection{Combination of the conditions}\label{subsec:combinationarguments}

In this section, we show in some detail how to combine the above conditions to rule out theories described by the action \eqref{GeneralAction}. Here we tackle rather simple cases in order to show in the clearest way how the requirements combine together. Part of this analysis was already performed in \cite{Ayuso:2014jda}. However, surprisingly, the authors have not drawn the same conclusion as us concerning the fate of these kinds of theories. To simplify the argument, in this subsection we consider the Horndeski form -- $i.e.$ the second-order equations of motion, instead of the DHOST form -- $i.e.$ the degeneracy condition, for the condition \ref{condition:2ndorder-eom}. 

\paragraph{Higher matter couplings} \label{subsec:higher-matter-general} We warm up by coming back to a case already discussed in the condition \ref{condition:matterismatter} of \cref{subsec:requirements}. We consider a Lagrangian of the form
\begin{align}
S&=\frac{M_P^2}2\int  \, d^4x \sqrt{g} \, \big\{ f(R) + f_1(\mathcal{L}_m) \big\} + \int d^4x \sqrt{-g} \, \mathcal{L}_m.  \label{only_matter}
\end{align} 
This case is almost trivial, yet it shows how to use the conditions of the previous section. Condition \ref{condition:notscalartensor} basically asks that the matter Lagrangian appearing as the argument of the function $f_1$ is the same as the one in the last part of \cref{only_matter}, namely true matter including the SM. Then one can trivially rewrite the action as:
\begin{align}
S&=\frac{M_P^2}2\int  \, d^4x \sqrt{g} \, f(R) + \int d^4x \sqrt{-g} \, \left\{\mathcal{L}_m + f_1(\mathcal{L}_m)\right\},  \label{only_matter_2}
\end{align} 
showing that the theory under study reduces to an $f(R)$ theory with a true matter Lagrangian given by:
\begin{equation}
\mathcal{L}_m^{(true)}=\mathcal{L}_m+f_1(\mathcal{L}_m).
\end{equation}
This has been noted already by \cite{Fisher:2019ekh}. Condition \ref{condition:matterismatter} then rules out this theory since it is not a proper modification of gravity. Moreover, for generic Lagrangian $\mathcal{L}_m$, if the energy scale of the modification is lower than the weak scale, the above action will be in contradiction with particle physics experiments, hence ruled out by condition \ref{condition:obsconstraints}.

We stress once again that in cosmological applications where such Lagrangians are used to describe our early Universe, hence when the matter Lagrangian does not describe today's Standard Model but $e.g.$ only a scalar field $\phi$, \cref{only_matter} only describe a scalar-tensor theory of gravity. Such theory is thus ruled out by condition \ref{condition:notscalartensor} in the sense that is not actually coupling matter with gravity.

\paragraph{Couplings between matter Lagrangian and scalar curvature}
\label{subsec:matter-curvature-general}
We now study less obvious cases which will highlight the main issue faced with theories described by \cref{GeneralAction}: they contain light ghosts. Here we give general arguments and show why theories of the form $f(R,\mathcal{L}_m)$ or $f(R_{\mu\nu}T^{\mu\nu})$ are not self-consistent. We study in more detail the case of theories of the form $f(R,T)$ in \cref{sec:specifics}.
We start with a Lagrangian of the form:
\begin{align}
S=\frac{M_P^2}2\int  \, d^4x \sqrt{g} \, f_1(R,\mathcal{L}_m)  + \int d^4x \sqrt{-g} \, \mathcal{L}_m, \label{couplingRLmatter}
\end{align}
and first consider the simple case
\begin{equation}
f_1(R,\mathcal{L}_m)=R \, h(\mathcal{L}_m), \label{specificCoupling}
\end{equation}
for an arbitrary function $h$.

From the conditions \ref{condition:matterismatter} and \ref{condition:obsconstraints}, we know that the matter sector should contain at least a scalar and a vector field and their associated kinetic terms; hence the matter Lagrangian should include:
\begin{align}
    \mathcal{L}_m &= - \frac 12 \nabla_{\mu}\phi \nabla^{\mu}\phi - m^2\phi^2 + \lambda \phi^4 - F^{\mu\nu}F_{\mu\nu} + \ldots \nonumber\\
    &= -X- m^2\phi^2 + \lambda \phi^4 - F^{\mu\nu}F_{\mu\nu} + \ldots, \label{kineticTerms}
\end{align}
where in the last equality we introduced $X\equiv\nabla_{\mu}\phi\nabla^{\mu}\phi$. The Lagrangian \eqref{couplingRLmatter} thus contains couplings between the Ricci scalar and derivative operators of a scalar and vector fields. 

Such couplings are however highly constrained. Indeed, as explained in condition \ref{condition:2ndorder-eom}, they lead in general to higher-derivative operators, indicating the presence of ghosts. The classification of higher-order couplings allowing for second-order equations of motions (more generally, fulfilling the degeneracy condition), hence preventing the presence of ghosts, has been performed by Horndeski \cite{Horndeski:1974wa,Horndeski:1976gi}. To avoid a ghost instability, the Lagrangian coupling the Ricci scalar $R$ with the scalar kinetic terms $X$ is required to take the form:
\begin{equation}
\mathcal{L}^H_4=G_4(\phi,X) R - 2 G_{4,X}(\phi,X)\left((\Box\phi)^2-\nabla^{\mu}\nabla^{\nu}\phi\nabla_{\mu}\nabla_{\nu}\phi\right),
\end{equation}
where $G_{4,X}$ denotes the derivative of $G_4(\phi,X)$ with respect to the variable $X$. In the case of the  Lagrangian \eqref{couplingRLmatter} with the simple coupling of \cref{specificCoupling}, due to the presence of the kinetic terms of the matter Lagrangian  \eqref{kineticTerms}, the function $G_4$ is identified to:
\begin{equation}
G_4(X,\phi)=h(-X-m^2\phi^2+\lambda \phi^4),
\end{equation}
so that the four-derivative terms for the scalar field in the full Lagrangian should exactly be:
\begin{equation}
\mathcal{L}_4 = 2 h'(-X-m^2\phi^2+\lambda \phi^4)\left((\Box\phi)^2-\nabla^{\mu}\nabla^{\nu}\phi\nabla_{\mu}\nabla_{\nu}\phi\right).
\end{equation}
The scalar sector of the matter Lagrangian should thus start exactly as:
\begin{equation}
\mathcal{L}_m^{\phi}= - \frac 12 \nabla_{\mu}\phi \nabla^{\mu}\phi - m^2\phi^2 + \lambda \phi^4 +  2 h'(-X-m^2\phi^2+\lambda \phi^4)\left((\Box\phi)^2-\nabla^{\mu}\nabla^{\nu}\phi\nabla_{\mu}\nabla_{\nu}\phi\right) + \ldots,
\end{equation}
where the scale of the higher-derivative terms is exactly determined by the gravity coupling function $h$ of \cref{specificCoupling}. Such terms, if not sufficiently suppressed, drastically modify the matter Lagrangian and should be detected by experiments. We thus expect that condition \ref{condition:obsconstraints} generally rules out such model. 

We now show that there is an even more important problem. Indeed, to allow for second-order equations of motion the couplings of the vector field to the curvature should be of the form \cite{Horndeski:1976gi}:
\begin{align}
 \mathcal{L}_V^H&= -\frac 12 \epsilon^{\mu\nu\rho\sigma}\epsilon^{\alpha\beta\gamma\delta} F_{\mu\nu}F_{\alpha\beta}R_{\rho\sigma\gamma\delta} \nonumber\\
 &=R F_{\mu\nu}F^{\mu\nu}-4 R_{\mu\nu}F^{\mu\sigma}F^{\nu}_{\,\,\sigma}+R_{\mu\nu\rho\sigma}F^{\mu\nu}F^{\rho\sigma}. \label{HcouplingV}
\end{align}
Hence, the fate of $f(R,\mathcal{L}_m)$ is worse than one could expect by looking only at the scalar sector. Indeed, since the last two terms of the above equation cannot be obtained from the Lagrangian \eqref{couplingRLmatter}, the presence of a ghost instability in the vector sector is unavoidable.

\paragraph{Couplings between matter stress-energy tensor and curvature}

One could try to solve this issue by considering a more general form of the Lagrangian, introducing couplings between the stress-energy tensor $T^{\mu\nu}$ and the Ricci curvature tensor.  For scalar and vector matter sectors the stress-energy tensor read:
\begin{equation}
T_{\mu\nu}= \nabla_{\mu}\phi\nabla_{\nu}\phi -g_{\mu\nu}\left(\frac 12 \nabla_{\mu}\phi\nabla^{\mu}\phi+m^2\phi^2-\lambda\phi^4\right) -F_{\mu\sigma}F_{\nu}^{\,\,\,\sigma}+ g_{\mu\nu}\frac{F^2}{4}, \label{Tmunuscalarvector}
\end{equation}
and we thus could consider a Lagrangian including a coupling of the form:
\begin{equation}
S=\frac{M_P^2}{2}\int d^4 x\sqrt{g} \, f_3(R_{\mu\nu}T^{\mu\nu}) + \int d^4 x \sqrt{-g}\mathcal{L}_m. \label{LRmunuTmunu}
\end{equation}
Here again the absence of a ghost, or more precisely, the requirement of second-order equations of motion (and more generally, of the degeneracy condition), as explained in Condition \ref{condition:2ndorder-eom}, highly constrains such coupling. For a scalar field, the only coupling with the Ricci curvature leading to second-order e.o.m must be of the form \cite{Horndeski:1974wa}:
\begin{equation}
\mathcal{L}^H_5=G_5(\phi,X)G_{\mu\nu}\nabla^{\mu}\nabla^{\nu}\phi+\frac{1}{6}G_{5,X}(\phi,X)\left[(\Box \phi)^3-3\Box \phi(\nabla_{\mu}\nabla_{\nu}\phi)^2+2(\nabla_{\mu}\nabla_{\mu}\phi)^3\right], \label{LH5}
\end{equation}
whereas for the vector field, the allowed coupling is included in \cref{HcouplingV}. 

In our case the absence of higher-derivative terms for $\phi$, as required by Condition \ref{condition:obsconstraints} for $e.g.$ the standard model sector, implied that the function $G_5$ in \cref{LH5} only contains $\phi$, such that the allowed form is:
\begin{align}
\mathcal{L}_5^H=G_5(\phi)G_{\mu\nu}\nabla^{\mu}\nabla^{\nu}\phi&=-G_{5,\phi}(\phi)G^{\mu\nu}\nabla_{\mu}\phi\nabla_{\nu}\phi + \rm{tot.}\nonumber\\
&=-G_{5,\phi}(\phi)\left(R^{\mu\nu}-\frac{1}{2} g^{\mu\nu} R \right) \nabla_{\mu}\phi\nabla_{\nu}\phi + \rm{tot.}\nonumber\\
&=-G_{5,\phi}(\phi)R^{\mu\nu} \left(\nabla_{\mu}\phi\nabla_{\nu}\phi -\frac{1}{2} g_{\mu\nu} \nabla^{\mu}\phi\nabla_{\mu}\phi\right) + \rm{tot.}
\end{align}
Using the expression \cref{Tmunuscalarvector} for the stress-energy tensor, this allowed term can be obtained using:
\begin{align}
-G_{5,\phi}(\phi) \left(R^{\mu\nu}T_{\mu\nu} + R \,[m^2\phi^2 -\lambda \phi^4] \right)= \mathcal{L}_5^H + \rm{vector \,\, part}.
\end{align}
The vector part induced by the above coupling is then of the form:
\begin{align}
-G_{5,\phi}(\phi) R^{\mu\nu}T_{\mu\nu} \supset -\frac{1}{4} G_{5,\phi}(\phi)\left( R \, F_{\mu\nu} F^{\mu\nu}-4 R_{\mu\nu}F^{\mu\sigma}F^{\nu}_{\,\,\sigma}\right).
\end{align}
Hence we see that the last term of \cref{HcouplingV} is again missing in order to obtain second-order equations of motion.

We thus conclude that, as soon as one considers a matter sector including both a scalar and a vector field, as is the case in the Standard Model of particle physics, one cannot couple the stress-energy tensor $T_{\mu\nu}$ to the Ricci curvature $R_{\mu\nu}$ and keep second order e.o.m. at the same time. Once again, we have combined conditions \ref{condition:matterismatter}, \ref{condition:obsconstraints}, and \ref{condition:2ndorder-eom} to rule out such coupling between matter and gravity.

\section{Finding the ghost in specific cases: \texorpdfstring{$f(R,T)$}{fRT} in low-curvature expansion}

\label{sec:specifics}
In section \ref{sec:generalarguments}, we have presented general arguments why $f(R,{\rm Matter})$ gravity theories have to be treated with care. We argued that, if not rewriting of GR with exotic matter or scalar-tensor gravity, they are not consistent due the presence of ghosts. In this section, we discuss $f(R,T)$ models in a low curvature expansion in greater detail and explicitly compute the scale at which the ghost field appears. 

To be more specific, we consider an action of the type
\begin{align}
    S &= \int d^4x \sqrt{-g}f(R,T) + \int d^4x \sqrt{-g} \, \mathcal{L}_m \nonumber \\
    &=\int d^4x \sqrt{-g}\left\{\vphantom{1^{6^2}}f_0(T)+Rf_1(T)+\ldots \right\} + \int d^4x \sqrt{-g} \, \mathcal{L}_m \label{Actionlow-curvExpan}
\end{align}
The first two terms of the expansion, written explicitly in the above equation, cover the most popular models in the literature. We deal with each of them first and then discuss the case when they are both considered at the same time. 

\subsection{\texorpdfstring{The simple case of $f_0(T)$}{Higher order matter couplings}} \label{subsec:higher-matter-example}

We start this section by studying the simple case where the gravitational part reduces to general relativity, hence we consider here that the function appearing in the second term of the expansion \eqref{Actionlow-curvExpan} is simply
\begin{equation}
f_1(T)\equiv\frac{M_P^2}{2},
\end{equation}
and are left only with the function $f_0(T)$. This case was already considered in \cite{Fisher:2019ekh}. The general arguments of section \ref{sec:generalarguments} apply here. We first consider the case of a linear function \eqref{linearf0} below, very popular in the literature, before turning to nonlinear forms of $f_0(T)$. 

\paragraph{Linear coupling function} We start our study by taking the linear function:
\begin{equation}
    f_0(T)\equiv \gamma_1 T, \label{linearf0}
\end{equation}
and consider first the trivial example of matter constituted of a free massless scalar. The matter Lagrangian and the trace of the stress-energy tensor thus read:
\begin{equation}
\mathcal{L}_m = -\frac{1}{2}\p^{\mu}\phi \p_{\mu}\phi, \qquad T=-\p^{\mu}\phi \p_{\mu}\phi, \label{freescalarLmT}
\end{equation}
so that the matter part of the action reads 
\begin{equation}
    S_m =- \int d^4 x \sqrt{-g}\left\{\frac{1+2\gamma_1}{2} \, \p^{\mu}\phi \p_{\mu}\phi\right\},
\end{equation}
which is nothing but the Lagrangian for a non-canonically normalized free massless scalar. Hence we see that the coupling \eqref{linearf0} leads to a redefinition of the matter sector, which in this trivial case is just a renormalization of the massless scalar.  

One can also include a scalar potential, turning to the study of a massive, interacting scalar field. The matter Lagrangian and trace $T$ read:
\begin{align}
    \mathcal{L}_m &=  -\frac{1}{2}\p^{\mu}\phi \p_{\mu}\phi -V(\phi)\equiv -\frac{1}{2}\p^{\mu}\phi \p_{\mu}\phi -\frac{m^2}{2}\phi^2 - \frac{\lambda}{4!} \phi^4, \\
    T &= -\p^{\mu}\phi \p_{\mu}\phi - 4 V(\phi).
\end{align}
The coupling \eqref{linearf0} thus leads to the following matter part of the action
\begin{equation}
    S_m = \int d^4x \sqrt{-g} \left\{-\frac{1+2\gamma_1}{2}\p^{\mu}\phi \p_{\mu}\phi-(1+4\gamma_1)V(\phi)\right\}.
\end{equation}

Here again, we see that the scalar is not canonically normalized and that its potential has been rescaled. To be fully explicit, we rewrite the full matter action after normalization: 
\begin{equation}
    S \supset \int d^4x \sqrt{-g} \left\{-\frac{1}{2}\p^{\mu}\phi \p_{\mu}\phi-\frac{(1+4\gamma_1)m^2}{2(1+2\gamma_1)^2}\phi^2-\frac{\lambda(1+4\gamma_1)}{(1+2\gamma_1)^2 4!}\phi^4\right\}.
\end{equation}
Hence, as advertised, even when the scalar potential is included, the additional term $f_0(T)$ only leads to a redefinition of the matter sector, namely the mass of the scalar and its self-interaction parameter. Such argument has already been raised in \cite{Fisher:2019ekh}. We discussed above the case of scalar matter but similar arguments naturally holds for fields of the matter sector with different spins.

\paragraph{Non-linear coupling function} We now consider a nonlinear function $f_0(T)$. We already had a similar discussion in \cref{subsec:matter-curvature-general} for a function of the matter Lagrangian $f_1(\mathcal{L}_m)$, which we ruled out following strictly the requirement of condition \ref{condition:matterismatter}. Here again, it is clear that such term introduces new interactions. In the case where $\mathcal{L}_m$ is the Lagrangian of a scalar, for example in inflationary models, $f_0(T)$ only leads to a new matter Lagrangian $\mathcal{L}^{(true)}$, rather than a modification of gravity, as discussed in \ref{sec:generalarguments}. In the case where $\mathcal{L}_m$ is really the Standard Model Lagrangian, such new interactions are additionally subjected to particle physics constraints, and the model is often also ruled out by condition \ref{condition:obsconstraints}.

We now analyze an explicit nonlinear example and, for full concreteness, show how the matter Lagrangian is modified to obtain a new ``true'' Lagrangian, which shall be subject to experimental constraints. We first expand $f_{0}(T)$ into 
\begin{equation}
    f_0(T)= \gamma_1 T + \frac{1}{2}\gamma_2 T^2+\ldots
\end{equation}
 According to our points \ref{condition:matterismatter} and \ref{condition:notscalartensor}, the matter Lagrangian $\mathcal{L}_m$  should contain the Standard Model and in particular the two massive $W$-boson terms, which read:
\begin{equation}
    \mathcal{L}_m \supset -\frac{1}{2}(\p_{\mu}W^+_{\nu}-\p_{\nu}W^+_{\mu})(\p^{\mu}W_-^{\nu}-\p^{\nu}W_-^{\mu})+\frac{m_W^2}{2}W^+_{\mu}W_-^{\mu}.
\end{equation}
The stress-energy tensor then contains 
\begin{equation}
    T^{\mu\nu} \supset T_{W}^{\mu \nu}\equiv -2 F^{\mu \beta} (F^{\nu}_{\beta})^{\dagger}+\frac{1}{2}g^{\mu \nu}F^{\alpha \beta} (F_{\alpha \beta})^{\dagger}-\frac{1}{2} g^{\mu \nu} m_W^2 W^+_{\alpha} W_-^{\alpha},
\end{equation}
where here $F^{\mu \beta} = \p^{\mu}W_+^{\beta}-\p^{\beta}W_+^{\mu}$.
The kinetic term drops out of the trace, such that we obtain:
\begin{equation}
    T_W=-2 m^2 W_{\alpha}^+ W^{\alpha}_-, \qquad 
     \frac{1}{2}\gamma_2 T_W^2 \supset 2\gamma_2 m_W^4 (W^+_{\alpha} W_-^{\alpha})^2. \label{W interaction modification}
\end{equation}
Notice that this is the only quartic term in $T^2$ involving the $W$ boson only, since the kinetic and mass terms above are the only quadratic terms in $T$ that contain the $W$ boson only. 

The quartic interaction of the $W$ boson in the Standard Model has the form\cite{Schwartz:2014sze}:
\begin{equation}\label{SM W quartic interaction}
    \mathcal{L}_{SM} \supset -\frac{g^2}{2}((W^+_{\alpha} W_-^{\alpha})^2-W^+_{\alpha} W_+^{\alpha} W^-_{\beta} W_-^{\beta}),
\end{equation}
so that the terms coming from $T^2_W$, shown in \eqref{W interaction modification}, cannot be absorbed into the Standard Model Lagrangian via a redefinition of fields and couplings, 
and would lead to new physics. In particular, they change the ratio of the two interaction channels in \eqref{SM W quartic interaction}. Already from \eqref{W interaction modification}, we see that partial wave unitarity demands (see, for example, \cite{Schwartz:2014sze}) that 
\begin{equation}
    |\gamma_2 m_W^4|\lesssim 1,
\end{equation}
or in other words, that the mass scale associated with new physics must be larger than the weak scale. This bound can be further sharpened by experimental results on $W$-boson scattering. In the notation of \cite{Baak:2013fwa}, the amplitude associated with \eqref{W interaction modification} is given by
\begin{equation}
    \mathcal{A}= 4g^2 \alpha_5.
\end{equation}
The magnitude of the dimensionless coefficient  $\alpha_5$ depends on the scale of new physics and is experimentally constrained \cite{ATLAS:2016snd} to lie in the $95\%$ confidence interval:
\begin{equation}
    \alpha_5 \in [-0.22,0.22] .
\end{equation}
In our case, the amplitude is equal to $8\gamma_2 m_W^4$, and so, using that $g^2 = 4\pi \alpha_{ew} /\sin^2(\theta_{W}) \approx 0.42$ \cite{ParticleDataGroup:2022pth} at the $m_Z$ scale, we obtain 
\begin{equation}
    |\gamma_2 m_W^4| \leq 0.05.
\end{equation}
This bound means that the coupling $1/2 \, \gamma_2 T^2$ is irrelevant for any process whose energy density is below $m_W^4$, hence the theory loses its interest for $e.g$ cosmology today, by virtue of the condition \ref{condition:energyscale}. In \cite{Flanagan:2003rb} the Palatini version of $f(R)$ gravity was excluded by a similar approach relying on the bounds from electron-electron scattering on higher-order matter interactions. 

We conclude this example by mentioning that for modifications of the theory that are only relevant at scales above the weak scale, there are no constraints from particle physics measurements today. However, once again, such modifications have to be seen as modifications of the matter sector only, not as a  modification of the theory of gravity, see condition \ref{condition:matterismatter}.

Moreover, we emphasize that, apart from being solely modifications of the matter sector, couplings of the type $f_0(T)$ violate the principle of effective field theories that permits arbitrary couplings in the absence of protecting symmetry. Indeed, terms such as $T^2$ introduce specific relations between the coefficients of operators of different dimensions, which are not supported by any {\it a priori} symmetry argument.

\subsection{\texorpdfstring{$Rf_1(T)$}{matter-curvature coupling} in Einstein frame for a free massless scalar}
\label{subsec:matter-curvature-example}
We now go to the next order in the low-curvature expansion of \cref{Actionlow-curvExpan}, which introduces a true coupling between matter and gravity. Our goal is to demonstrate explicitly the appearance of a ghost related to this coupling and to determine its energy scale.

We thus consider the following Jordan frame action:
\begin{equation}
    S = \frac{M_P^2}{2} \int d^4x \sqrt{-g} R f_1(T)+ \int d^4x \sqrt{-g} \, \mathcal{L}_m(g_{\mu \nu}, \chi, \partial_{\mu} \chi), \label{Actionf1Lm}
\end{equation}
where $\chi$ is a shorthand notation for the matter fields. We perform the Weyl transformation  $g_{\mu\nu}=1/f_1(T)\Tilde{g}_{\mu\nu}$ to go to the Einstein frame, in which the action reads
\begin{equation}\label{f(T)R action- Einstein frame}
    S= \frac{M_P^2}{2} \int d^4x \sqrt{-\Tilde{g}} \left\{ \Tilde{R}-\frac{3}{2f_1(T)^2}\p^{\mu}f_1(T)\p_{\mu}f_1(T) \right\} +\int d^4 x \frac{\sqrt{-\tilde{g}}}{f_1(T)^2}\, \mathcal{L}_m\big(\frac{\tilde{g}_{\mu \nu}}{f_1(T)}, \chi, \p_{\mu} \chi \big). 
\end{equation}
We comment on the fact that here $T=T(\chi, \p\chi)$ is not a field variable but rather a composite field, namely a function of all the matter fields and their derivatives. Hence, the second term in \eqref{f(T)R action- Einstein frame} leads to nonlinear second-time derivative terms, since $T$ already contains first-time derivatives from the standard kinetic terms of the matter fields $\chi$. This is the standard hint for the appearance of a ghost.

\paragraph{Some assumptions to simplify our study}
 To demonstrate the appearance of a ghost, we now take specific forms of coupling function $f_1(T)$ and matter Lagrangian. We  choose the coupling: 
\begin{equation}
f_1(T)\equiv 1+\lambda_1 T, \label{f1T}
\end{equation}
 so that the action \eqref{Actionf1Lm} describes Einstein gravity together with the simplest matter-curvature coupling. In the following, we will work in the regime where this coupling is small with respect to the standard Einstein-Hilbert term:
\begin{equation}
\text{\it regime under study:} \quad \xi \equiv \lambda_1 T \ll 1. \label{definxi}
\end{equation}
In this regime, \cref{f1T} can be seen as the lower terms of the $T$ expansion of a more general coupling $f_1(T)$. We will show that even the regime with small non-vanishing $\xi$ is sufficient to show the appearance of a ghost. It is thus expected that going to larger $\xi$ only worsens this problem.  From \cref{f1T}, $\lambda_1$ has a dimension of $[\lambda_1]=M^{-4}$.

 We choose to consider matter made of a massless free scalar $\phi$, the Lagrangian and stress-energy tensor of which is given in the Jordan frame by \cref{freescalarLmT}. They can be expressed using the Einstein frame metric of \cref{f(T)R action- Einstein frame} as:
 \begin{align}
\mathcal{L}_m &= -\frac{1}{2}g^{\mu\nu}\p_{\nu}\phi \p_{\mu}\phi=-\frac{f_1(T)}{2}\tilde{g}^{\mu\nu}\p_{\nu}\phi \p_{\mu}\phi = -\frac{1+\lambda_1 T}{2}\tilde{g}^{\mu\nu}\p_{\nu}\phi\p_{\mu}\phi, \\
T&=2 \mathcal{L}_m=-(1+\lambda_1 T)\tilde{g}^{\mu\nu}\p_{\nu}\phi \p_{\mu}\phi, \label{TEinsteinmetric}
 \end{align}
 where we used the explicit expression for $f_1(T)$ given in \cref{f1T}. We stress that the choice of matter made of a free massless scalar is only a matter of simplicity and brevity of expressions. The result of the following arguments does not change qualitatively when considering more complicated matter actions as the one of the SM.

We can now invert \cref{TEinsteinmetric} to write:
\begin{equation}
    1+\lambda_1 T = \frac{1}{1+\lambda_1 \p_{\mu} \phi \Tilde{g}^{\mu \nu} \p_{\nu} \phi},
\end{equation}
and express \eqref{f(T)R action- Einstein frame} only in terms of the scalar field function $X\equiv-\p_{\mu}\phi \Tilde{g}^{\mu \nu} \p_{\nu}\phi$. The action then reads:
\begin{align}\label{action-free massless scalar}
    S = & \frac{M_P^2}{2}\int d^4x  \sqrt{-\Tilde{g}} \left\{\Tilde{R}-\frac{3}{2}\frac{\lambda_1^2}{(1-\lambda_1 X)^2} \p_{\alpha}X \Tilde{g}^{\alpha \beta} \p_{\beta}X \right\} + \int d^4x \sqrt{-\tilde g} \frac{1}{2}(1-\lambda_1 X)X \nonumber \\
     = & \frac{M_P^2}{2}\int d^4x  \sqrt{-\Tilde{g}} \left\{\Tilde{R}-\frac{6 \lambda_1^2 }{(1+\lambda_1 \p_{\mu}\phi \Tilde{g}^{\mu \nu} \p_{\nu}\phi)^2} (\nabla^{\alpha}\p^{\mu}\phi) \p_{\mu}\phi (\nabla_{\alpha}\p^{\nu}\phi) \p_{\nu}\phi \nonumber\right\} \\
     & - \int d^4x \sqrt{-\tilde g} \frac{1}{2}(1+\lambda_1 \p^{\mu}\phi \, \p_{\mu}\phi)\p^{\nu}\phi \, \p_{\nu}\phi, 
\end{align}
where $\p^{\mu}=\tilde{g}^{\mu\nu}\p_{\nu}$, $\nabla^{\mu}=\tilde{g}^{\mu\nu}\nabla_{\nu}$, and $\nabla_{\nu}$ is the covariant derivative compatible with $\tilde{g}_{\mu\nu}$.

\paragraph{Not Horndeski and not DHOST} \label{subsubsec:not-DHOST}

As already mentioned, the second derivative terms present in the action \eqref{action-free massless scalar} generically yield higher-order equations of motion signaling the presence of ghosts. The only exceptions are for a Lagrangian of the Horndeski \cite{Horndeski:1974wa} type, keeping second order e.o.m, or a degenerate Lagrangian (DHOST) which remains ghost-free albeit having higher-order e.o.m. \cite{Langlois:2015cwa}. 

One can repeat the study done in \cref{subsec:matter-curvature-general} below \cref{couplingRLmatter} to show that, as the action \eqref{action-free massless scalar} is not of the Horndeski type, the equations of motion are not second-order. 

The authors of \cite{Langlois:2015cwa} have listed the degenerate conditions the action of generic scalar-tensor theories must satisfy to be free from ghost instabilities. Using their notations, the action \eqref{action-free massless scalar} can be written as
\begin{equation}
    S = \int d^4x \sqrt{-\tilde{g}}\left\{\frac{M_P^2}{2} \tilde{R} + C^{\mu \nu, \rho \sigma}\nabla_{\mu}\p_{\nu}\phi\nabla_{\rho}\p_{\sigma} \phi- \frac{1}{2}(1+\lambda_1 \p_{\mu}\phi \Tilde{g}^{\mu \nu} \p_{\nu}\phi)\p_{\mu}\phi \Tilde{g}^{\mu \nu} \p_{\nu}\phi \right\},
\end{equation}
where we defined the function:
\begin{equation}
    C^{\mu\nu,\rho\sigma}\equiv \frac {3  M_P^2\lambda_1^2  }{4(1+\lambda_1 \p_{\mu}\phi \Tilde{g}^{\mu \nu} \p_{\nu}\phi)^2}  \left(\partial^{\mu}\phi\p^{\rho}\phi \tilde{g}^{\nu\sigma}+ \partial^{\nu}\phi\p^{\rho}\phi \tilde{g}^{\mu\sigma}+\partial^{\mu}\phi\p^{\sigma}\phi \tilde{g}^{\nu\rho}+\partial^{\nu}\phi\p^{\sigma}\phi \tilde{g}^{\mu\rho} \right).
\end{equation}
The $\alpha_i$ coefficients defined in \cite{Langlois:2015cwa} to express the degenerate conditions then read 
\begin{equation}
   \alpha_1 = \alpha_2 = \alpha_3 = \alpha_5 = 0, \qquad \alpha_4 = \frac{3M_P^2 \lambda_1^2}{(1+\lambda_1 \p_{\mu}\phi \Tilde{g}^{\mu \nu} \p_{\nu}\phi)^2},
\end{equation}
while (in their notation) $f\equiv M_P^2/2$ as we are in the Einstein frame. Without a dynamical metric, the degeneracy condition $\alpha_3 + \alpha_4 = 0$ is violated and the kinetic matrix is not degenerate. With dynamical metric, since $\alpha_1 + \alpha_2 =0$, the degeneracy condition $5.1$ of \cite{Langlois:2015cwa} implies $\alpha_4 =0$. Again, the degeneracy condition is not fulfilled and the theory has a ghost. 

\paragraph{Light ghost in a matter background}
\label{subsubsec:background-ghost}
The appearance of a ghost is not necessarily a problem as long as it appears at energies sufficiently higher than the scale at which the modified theory is used. Namely, it shall not appear in the effective theory below a cutoff scale that is sufficiently higher than the scale of applications of the theory. In this section, we show that for the action \eqref{action-free massless scalar}, this is not the case: the ghost appears at a scale depending on the parameter $\xi$ which is smaller than the scales relevant for cosmology, unless the value of $\xi$ is so small that we recover GR. We recall that as noted below \cref{f1T}, $\lambda_1$ has a dimension of inverse quarted mass $[\lambda_1]=M^{-4}$.

Although the action \eqref{action-free massless scalar} contains higher derivatives, they do not appear in terms quadratic in fields. Indeed, they come from the term $\partial_{\mu}X \partial^{\mu}X$, which is quartic in $\phi$. Therefore, the dispersion relation of the scalar $\phi$, obtained from the equation of motions, does not show the presence of a ghost on an empty background. The presence of the ghost is however clear when the scalar takes a background expectation value, as we show below. 

We will introduce explicitly a new variable accounting for the degree of freedom coming from higher-derivative terms. To do so,  we will only consider the scalar part of the action \cref{action-free massless scalar}. Indeed, we want to show the presence of a dangerous ghost mode, with momentum higher than the background but lower than the EFT cutoff. Hence we can neglect metric variations and we will work in a flat background. One could perform the same analysis in another background or without performing a background+perturbation analysis, and the conclusions relative to the presence of a ghost should be the same. We thus consider the Lagrangian:
\begin{align}
    \mathcal{L}= -\frac{3}{4}\frac{\lambda_1^2 M_P^2}{(1+\lambda_1\partial_{\nu} \phi \partial^{\nu} \phi )^2}\partial^{\mu} (\partial_{\nu} \phi \partial^{\nu} \phi)
    \partial_{\mu} (\partial_{\alpha} \phi \partial^{\alpha} \phi) - \lambda_1 \frac{(\partial_{\mu} \phi \partial^{\mu} \phi)^2}{2} - \frac{1}{2}\partial_{\mu} \phi \partial^{\mu} \phi.
\end{align}
We define a new field $\psi$ and enforce the condition $\psi = -\lambda_1 M_P \p^{\mu} \phi \p_{\mu} \phi$ through a Lagrange multiplier $\eta$. The part of the Lagrangian \eqref{action-free massless scalar} involving the scalar field is thus equivalent to:
\begin{equation}
    \mathcal{L}=-\frac{3}{4}\frac{1}{(1-\psi/M_P)^2}\p^{\mu} \psi \p_{\mu} \psi - \frac{1}{2}\p_{\mu} \phi \p^{\mu} \phi - \frac{\psi^2}{2\lambda_1M_P^2} + \eta \, \big(\frac{\psi}{M_P}+\lambda_1 \p^{\mu} \phi \p_{\mu} \phi\big).
\end{equation}
We now introduce a background expectation value and perturbation for each of the fields of the above Lagrangian through:
\begin{align}
 \psi \equiv \psi_{(0)} + \psi_{(1)}, \qquad \phi \equiv \phi_{(0)} +  \phi_{(1)}, \qquad \eta \equiv \eta_{(0)} + \eta_{(1)} .
\end{align}
The background equations of motion are:
\begin{align}
    &\eta_{(0)} - \frac{\psi_{(0)}}{M_P\lambda_1}+\frac{3}{2(1-\psi_{(0)}/M_P)^3}\p_{\mu}\psi_{(0)}\p^{\mu}\psi_{(0)}+\frac{3 M_P}{2(1-\psi_{(0)}/M_P)^2} \Box\psi_{(0)}=0, \label{alphabackground} \\
    & \Box \phi_{(0)} -2 \lambda_1 \p_{\mu} \big(\eta_{(0)} \p^{\mu}\phi_{(0)}\big) = 0,\\
    & \psi_{(0)} +\lambda_1 {M_P} \, \p^{\mu}\phi_{(0)} \p_{\mu}\phi_{(0)} = 0, \label{eometa0}
\end{align}
and the Lagrangian quadratic in the perturbations reads
\begin{align}
    \mathcal{L}^{(2)}= & -\frac{3}{4} \frac{1}{(1-\psi_{(0)}/M_P)^2} \p^{\mu}\psi_{(1)} \p_{\mu}\psi_{(1)} -  \frac{1-2\eta_{(0)} \lambda_1}{2}\p_{\mu}\phi_{(1)} \p^{\mu}\phi_{(1)} + 2 \lambda_1 \eta_{(1)} \p_{\mu}\phi_{(0)} \p^{\mu}\phi_{(1)} \nonumber\\
    & - \frac{\psi_{(1)}^2}{2\lambda_1M_P^2} + \frac{1}{M_P}\eta_{(1)} \psi_{(1)}. \label{L2order}
\end{align}

In \eqref{L2order} we have assumed that the background variation is much smaller than the perturbation one, hence we used ${\p_{\mu} \psi_{(1)}}/{\psi_{(1)}} \gg {\p_{\mu} \psi_{(0)}}/{\psi_{(0)}}$ to keep dominant terms. This assumption boils down to the fact that we study possible UV instabilities on a background of which the energy/momentum scales are below the cutoff of the theory.

When assuming that $\Box \psi_{(0)}/\psi_{(0)} \ll \frac{1}{\lambda_1 M_P^2}$, the background equation of motion for $\psi_{(0)}$ showed in \cref{alphabackground} leads to
\begin{equation}
    \eta_{(0)} = \frac{\psi_{(0)}}{\lambda_1 M_P}.
\end{equation}
The above assumption states that the background variations are smaller than the perturbation ones, as can be seen explicitly from  \cref{omega1,omega2}. Reinserting this relation for $\eta_{(0)}$ into the quadratic action gives
\begin{align}
    \mathcal{L}^{(2)}= -&\frac{3}{4}\frac{1}{(1-\psi_{(0)}/M_P)^2 }\partial_{\mu}\psi_{(1)} \partial^{\mu}\psi_{(1)} - \frac{\psi_{(1)}^2}{2\lambda_1 M_P^2} - \frac{1}{2}\left(1-2\frac{\psi_{(0)}}{M_P}\right)\partial_{\mu} \phi_{(1)} \partial^{\mu} \phi_{(1)}\nonumber \\ + &\eta_{(1)} \left( \frac{\psi_{(1)}}{M_P} + 2 \lambda_1 \partial^{\mu}\phi_{(1)} \partial_{\mu}\phi_{(0)} \right). \label{2ndorderL}
\end{align}

\paragraph{Adding yet an additional auxiliary field} To simplify our expressions, we first introduce the following notations for the background quantities:
\begin{align} \label{alphabetagamma}
    \alpha=\frac{3}{2(1-\psi_{(0)}/M_P)^2}, \quad \beta=1-2\psi_{(0)}/M_P, \quad \gamma=\frac{1}{\lambda_1 M_P^2}, \quad n^{\mu} = 2M_P \lambda_1 \partial^{\mu}\phi_{(0)}.
\end{align}
Note that from these definitions and the $\psi_{(0)}$ equation of motion \eqref{eometa0} we get the on-shell relation:
\begin{equation}
1-\beta= \frac{2 \psi_{(0)}}{M_P} = - \frac{1}{2} \gamma \,n^{\mu}n_{\mu}. \label{gamma n relation}
\end{equation}
The second order perturbation Lagrangian \eqref{2ndorderL} then becomes
\begin{equation}
    \mathcal{L} = -\frac{\alpha}{2}\partial_{\mu}\psi_{(1)} \partial^{\mu}\psi_{(1)} - \frac{\beta}{2} \partial_{\mu} \phi_{(1)} \partial^{\mu} \phi_{(1)} - \frac{\gamma}{2}\psi_{(1)}^2 + \frac{\eta_{(1)}}{M_P}(\psi_{(1)} + n^{\mu} \partial_{\mu} \phi_{(1)}).
\end{equation}
We first introduce a Lagrange multiplier $\lambda^{\mu}$ that replaces $\partial_{\mu}\psi_{(1)}$ with $P_{\mu}$ and rewrite into the following equivalent Lagrangian:
\begin{equation}
    \mathcal{L}_1 = -\frac{\alpha}{2}P_{\mu} P^{\mu} - \frac{\beta}{2} \partial_{\mu} \phi_{(1)} \partial^{\mu} \phi_{(1)} - \frac{\gamma}{2}\psi_{(1)}^2 + \frac{\eta_{(1)}}{M_P}(\psi_{(1)} + n^{\mu} \partial_{\mu} \phi_{(1)}) + (\lambda^{\mu} P_{\mu}+ \psi_{(1)} \partial_{\mu}\lambda^{\mu}).
\end{equation}
The equations of motion for ${\eta_{(1)}}$ and $P^{\mu}$ lead to:
\begin{equation}
\psi_{(1)}+n^{\mu}\p_{\mu}\phi_{(1)}=0, \qquad \lambda^{\mu}-\alpha P^{\mu}=0,
    \end{equation}
so that using them back in the Lagrangian leads to:
\begin{equation}
    \mathcal{L}_2 = \frac{1}{2\alpha}\lambda^{\mu} \lambda_{\mu} - \frac{\beta}{2} \partial_{\mu} \phi_{(1)} \partial^{\mu} \phi_{(1)} - \frac{\gamma}{2}(n^{\mu} \partial_{\mu} \phi_{(1)})^2 - n^{\mu} \partial_{\mu} \phi_{(1)} \partial_{\nu}\lambda^{\nu}.
\end{equation}
This Lagrangian can be written in matrix notations as
\begin{equation} \label{LagrangianbeforeLambdaint}
    \mathcal{L}_2 = -\frac{1}{2} 
      \partial_{\mu} \begin{pmatrix}
        \phi_{(1)} &  \lambda^{\alpha} \\
    \end{pmatrix}
\mathcal{G}^{\mu\nu}
    \,  \partial_{\nu} \begin{pmatrix}
         \phi_{(1)} \\
         \lambda^{\beta}
    \end{pmatrix} - \frac{1}{2} 
    \begin{pmatrix}
        \phi_{(1)} & \lambda^{\alpha} \\
    \end{pmatrix}
\mathcal{M}
    \begin{pmatrix}
        \phi_{(1)} \\ \lambda^{\beta} 
    \end{pmatrix},
\end{equation}
where the kinetic and mass matrices were defined as
\begin{equation}
    \mathcal{G}^{\mu\nu}=\begin{pmatrix}
        \beta g^{\mu \nu} + \gamma n^{\mu} n^{\nu} & n^{\mu} \delta^{\nu}_{\beta} \\
        n^{\nu} \delta^{\mu}_{\alpha}& 0 \\ 
    \end{pmatrix}, \qquad \mathcal{M}=    \begin{pmatrix}
        0 & 0 \\
        0 & -\frac{1}{\alpha} g_{\alpha \beta} \\ 
    \end{pmatrix}.
    \end{equation}
The kinetic matrix $\mathcal{G}^{\mu\nu}$ has zero eigenvalues, indicating the presence of non-dynamical fields. To show the presence of the ghost, we should integrate the non-dynamical fields and consider the kinetic matrix of the dynamical fields only. The latter shall not contain any zero eigenvalues and the presence of negative eigenvalues is equivalent to the presence of a ghost. 

\paragraph{Time-like background}
We now show how to proceed in the case of a time-like background, which is relevant for cosmological applications of the $f(R,T)$ models. We quickly comment on the fact that space-like backgrounds can possibly suffer from instantaneous modes or anisotropies not seen in experiments.
In a time-like background, we can choose coordinates such that $n^i=0$ for all spatial indices. After introducing: 
\begin{equation}
    C\equiv n^0=-2M_P \lambda_1 \p_t \phi_{(0)},
\end{equation}
the Lagrangian becomes
\begin{equation}
    \mathcal{L}_2= -\frac{1}{2\alpha} (\lambda^0)^2 + \frac{1}{2\alpha} (\lambda^i)^2 + \frac{\beta}{2} \dot{\phi}_{(1)}^2 - \frac{\beta}{2}(\partial_i \phi_{(1)})^2-\frac{\gamma C^2}{2} \dot{\phi}_{(1)}^2 - C \dot{\phi}_{(1)}\dot{(\lambda^0)} - C\dot{\phi}_{(1)} \partial_i \lambda^i.
\end{equation}
The $\lambda^i$ fields have no time derivative and can be integrated out. After integrating the last term by parts, and neglecting again the background derivatives, the equation of motion for $\lambda^i$ gives
\begin{equation}
    \lambda_i + \alpha C \partial_i \dot{\phi}_{(1)}=0.
\end{equation}
Reinserting into the Lagrangian then leads to
\begin{align}
    \mathcal{L}_2&= -\frac{1}{2\alpha} (\lambda^0)^2 + \frac{\beta}{2}\dot{\phi}_{(1)}^2 - \frac{\beta}{2}(\partial_i \phi_{(1)})^2 - \frac{\gamma C^2}{2} \dot{\phi}^2 - C \dot{\phi} \dot{(\lambda^0)} -\frac{\alpha C^2}{2} (\partial_i \dot{\phi}_{(1)})^2\nonumber\\
    &= \frac{1}{2} \begin{pmatrix}
        \dot{\phi}_{(1)} &  \dot{\lambda}^0 \\
    \end{pmatrix}
\mathcal{K}
    \, \begin{pmatrix}
        \dot{\phi}_{(1)} \\
         \dot{\lambda^0}
    \end{pmatrix} - \frac{1}{2} 
    \begin{pmatrix}
        \phi_{(1)} & \lambda^{0} \\
    \end{pmatrix}
\mathcal{M}
    \begin{pmatrix}
        \phi_{(1)} \\ \lambda^{0} 
    \end{pmatrix},
    \label{lagrangianlambdaintegrated}
\end{align}
where the kinetic and mass matrices read:
\begin{equation}
   \mathcal{K}= \begin{pmatrix}
        \beta -\gamma C^2 - \alpha C^2 \Vec{k}^2 & - C \\
        -C & 0 \\
    \end{pmatrix},
    \qquad   \mathcal{M}= \begin{pmatrix}
        \beta \vec{k}^2  & 0\\
        0 & \frac{1}{\alpha} \\
    \end{pmatrix}. \label{kinandmass}
\end{equation}
As the determinant of the kinetic matrix is negative, one of the eigenvalues is negative and the presence of the ghost is made explicit. We see that in the limit $C\rightarrow 0$, which corresponds to taking the background to zero, the kinetic matrix gets a zero eigenvalue, signaling that the extra degree of freedom becomes strongly coupled.

We now evaluate the scale at which the ghost appears through the dispersion relations of the different modes of the above theory. The equations of motion in Fourier space derived from the Lagrangian  \eqref{lagrangianlambdaintegrated} just read:
\begin{equation} 
   (\omega^2 \mathcal{K} - \mathcal{M})
    \begin{pmatrix}
        \phi_{(1)} \\
        \lambda^{0} \\
    \end{pmatrix} =  - \begin{pmatrix}
        \beta k^2 +\gamma C^2 \omega^2 + \alpha C^2 \omega^2 \Vec{k}^2 & \,\, C\omega^2 \\
        C \omega^2 & \,\, \frac{1}{\alpha} \\
    \end{pmatrix}  \begin{pmatrix}
        \phi_{(1)} \\
        \lambda^{0} \\
    \end{pmatrix} = 0.  \label{eommatrix2}
\end{equation}
Non-trivial solutions to the equations of motion exist when the above matrix is degenerate. Setting the determinant of that matrix to zero gives the condition
\begin{align}
 \beta k^2+ \gamma C^2  \omega ^2+ \alpha C^2 \omega^2 k^2 =0, \label{disprel1}
\end{align}
and expressing $C^2=2(1-\beta)/\gamma$ using \cref{gamma n relation}, we  solve the dispersion relation and find the two solutions:
\begin{equation}
    \omega_{1,2}^2= \frac{2 \alpha (1-\beta) \Vec{k}^2-3\beta\gamma+2\gamma \pm \sqrt{(2 \alpha (1-\beta) \Vec{k}^2  - 3\beta\gamma + 2\gamma )^2 + 8 \alpha \beta \gamma (1-\beta) \Vec{k}^2}}{4 \alpha (1-\beta)}.
\end{equation} 
At large spatial momentum $\Vec{k}^2 \gg \gamma/(\alpha(1-\beta))$, the two solutions approximately become:
\begin{align}
  \omega_1^2 &\sim \Vec{k}^2 + \frac{\gamma}{\alpha}\sim \Vec{k}^2 + \frac{2(1-\psi_{(0)}/M_P)^2}{3\lambda_1 M_P^2}\sim   \Vec{k}^2 + \frac{2}{3\lambda_1 M_P^2}, \label{omega1}\\
\omega_2^2 &\sim -\frac{\beta \gamma }{2(1-\beta)\alpha} \left( 1- \frac{\gamma}{\alpha \Vec{k^2}} \right)\sim-\frac{1}{6 \dot{\phi}_{(0)}^2 \lambda_1^2 M_P^2} \Big( 1- \frac{2}{3\lambda_1 M_P^2 \Vec{k^2}} \Big).  \label{omega2}
\end{align}
We have used that $\psi_{(0)}/M_P=\lambda_1 \dot{\phi}_{(0)}^2 \ll 1$ to simplify the final expressions. 

We just found the dispersion relations related to the two different modes of the second-order Lagrangian. Previously we have also demonstrated the presence of a ghost by evaluating the sign of the determinant of the kinetic matrix. We will now determine the eigenmodes of the system and look at their kinetic term to determine which mode is the ghost mode. To calculate the eigenmodes it is convenient to express the system in terms of $C \lambda^0$ instead of $\lambda^0$, such that the new variable has the same mass dimension as $\phi_{(1)}$. The equations of motion \eqref{eommatrix2}, evaluated at the frequencies $\omega_{1,2}$ solutions of \cref{disprel1}, read
\begin{equation}
    \begin{pmatrix}
         \alpha C^2 \omega_{1,2}^4  & \,\, \omega_{1,2}^2 \\
     \omega_{1,2}^2 & \,\, \frac{1}{\alpha C^2} \\
    \end{pmatrix}
    \begin{pmatrix}
        \phi_{(1)} \\
        C \lambda^0 \\
    \end{pmatrix} = 0,
\end{equation}
so that the kernel is generated by:
\begin{equation}\label{eigenvector}
    \begin{pmatrix}
        1 \\
        - \alpha C^2 \omega_{1,2}^2 \\
    \end{pmatrix}.
\end{equation}
We can now calculate the expectation values of the kinetic matrix $\mathcal{K}$ of \cref{kinandmass}:
\begin{equation}
    \langle \mathcal{K} \rangle_{1,2} =  \frac{\beta \Vec{k}^2}{\omega_{1,2}^2} + \alpha C^2 \omega_{1,2}^2 .
\end{equation}
At large spatial momenta, the frequencies of \cref{omega1,omega2} can be approximated by $\omega_{1}^2 \approx \Vec{k}^2$ and $\omega_2^2 \approx - 2/(3C^2)$ so that the kinetic term expectation values read
\begin{equation}
    \langle \mathcal{K} \rangle_{1} \approx  \beta + \alpha C^2 \Vec{k}^2, \qquad
    \langle \mathcal{K} \rangle_{2}\approx - \frac{3}{2} C^2 \beta \Vec{k}^2 - \frac{2}{3}\alpha.
\end{equation}
As in our approximation $\beta = 1- {2\psi_{(0)}}/{M_P}\approx 1$, the first is positive while the second is negative, thus identifying the ghost with the second excitation.

The scale at which the ghost appears is thus given through \cref{omega2} as:
\begin{equation}
\Lambda_{gh}^2 \sim |\omega_2^2 |\sim\frac{1}{6 \lambda_1 \dot{\phi_{(0)}^2 }}   \frac{1}{\lambda_1 M_P^2}\sim \frac{1}{\xi} \frac{1}{\lambda_1 M_P^2} \lesssim \frac{H^2}{\xi^2}. \label{boundLambdaghost}
\end{equation}
Here we have expressed the background energy density $\dot{\phi}_{(0)}^2$ in terms of $\xi=\lambda_1 T$ defined in \cref{definxi}. The last inequality is obtained using that the background energy density for $\phi$ is smaller than the total energy density. The latter being related to the Hubble scale, this thus corresponds to the bound $|T| \lesssim \rho_{tot} \sim H^2 M_P^2$.

{\it \underline{Late time cosmology:}} It is now clear that if the model is applied to model building at the present time, for instance, to describe dark energy, the bound \eqref{boundLambdaghost} shows that a ghost appears at scales lower than $H_0^2/\xi^2$. Due to the smallness of $H_0$, the ghost scale will be lower than the electroweak scale, and thus in conflict with particle physics, as long as $\xi \gtrsim H_0/m_W \sim 10^{-44}$. On the other hand, we do not expect models with $\xi\lesssim 10^{-44}$ to differ quantitatively from GR and thus be good models to explain new physics today.

{\it \underline{Early universe cosmology:}} If such models are rather used to describe early Universe physics, close to the end of the inflationary phase, the bound \eqref{boundLambdaghost} implies that $\Lambda_{gh} \lesssim H_{inf}/\xi$. It might seem that for small $\xi$ there is room for $\Lambda_{gh}>H_{inf}$, so that physics at the horizon scale does not excite the ghosts. However, only a few e-folds $N\sim \log(\xi)$ before, the horizon scale modes would have been inside the horizon and ghostly, ruining the description of perturbations during inflation. Stated otherwise, this model can at best be used only during a period of $N\sim \log(\xi)$ e-foldings, which except for excessively small $\xi$ is in contradiction with the minimal required period of inflation. Here again, in the case where $\xi$ is excessively small, we expect to recover GR so that the model is useless for the purpose of modifying gravity in early universe. 

\paragraph{Low cutoff from the EFT perspective}
\label{subsubsec:cutoff-from-EFT-Hedron}
In this paragraph, we offer a complementary perspective on the study of the inconsistency of the Lagrangian \eqref{action-free massless scalar}. Seen as low energy approximation, it naturally enters the framework of effective field theories (EFT). The consistency of the latter have undergone great scrutiny in the past years, with or without coupling to gravity, see $e.g.$ \cite{Adams:2006sv,Bellazzini:2014waa,Nicolis:2009qm,Baumann:2015nta,Cheung:2016yqr,deRham:2017avq,deRham:2017zjm,deRham:2018qqo}. 

Even if not necessary from the low-energy point of view, one can wonder what requirements the theory described by the Lagrangian \eqref{action-free massless scalar} must fulfill in order to have a unitary, causal, Lorentz-invariant, and positive energy UV-completion. S-matrix arguments then place constraints on the low-energy effective couplings \cite{Aharonov:1969vu,Pham:1985cr,Adams:2006sv}. Describing and constraining modified gravity models using effective field theory techniques is not a novel approach, see $e.g.$ \cite{Gubitosi:2012hu,Bloomfield:2012ff,Gleyzes:2013ooa,Gleyzes:2014rba,Bellini:2014fua,Kase:2014cwa,DeFelice:2015isa,Langlois:2017mxy,Frusciante:2019xia,Renevey:2020tvr,Lagos:2016wyv,Lagos:2017hdr,deRham:2021fpu}. 

The bounds are formulated using the Wilson coefficients of the $2\rightarrow 2$ amplitude as they are unambiguous, in contrast to couplings in the Lagrangian, which can for instance change under integration by parts. Moreover, as they rely on scattering amplitudes only, the bounds shown in this section can in principle be generalized to more complicated contents of the energy-momentum tensor compared to the simple case of a scalar field considered in the previous paragraphs. 

The tree-level $2\rightarrow 2$ scattering amplitude of \eqref{action-free massless scalar} is given by
\begin{align}
    \mathcal{A}_{2\rightarrow 2}= & \frac{3}{2}M_P^2 \lambda_1^2\left[-3s^2 t -3 t^2 s\right] - \lambda_1 \left[2s^2+2t^2+2st\right]\\
    = & \frac{3}{2}M_P^2 \lambda_1^2\left[-3\left(s+\frac{t}{2}\right)^2 t^1-\frac{3}{4}t^3\right]-2\lambda_1 \left[\left(s+\frac{t}{2}\right)^2 t^0+ \frac{3}{4}\left(s+\frac{t}{2}\right)^0 t^2\right] \nonumber,
\end{align}
where we have defined the Mandelstam variables in the usual form  $s=-(p_1+p_2)^2$, $t=-(p_1-p_3)^2$, $u=-(p_1-p_4)^2$. The Wilson coefficients $g_{k,q}$, extracted from the expansion 
\begin{equation}
    \mathcal{A}_{2\rightarrow 2} = \sum_{q,k} g_{k,q}\left(s+\frac{t}{2}\right)^{k-q} t^q,
\end{equation}
thus read:
\begin{equation}
g_{2,0}=-2\lambda_1, \quad g_{2,2}= - 3\lambda_1/2, \quad g_{3,1}=-9 M_P^2 \lambda_1^2/2, \quad g_{3,3}=-9M_P^2 \lambda_1^2/8 . \label{Wilsoncoeffs}
\end{equation}
According to the recent work \cite{Chiang:2021ziz}, if the theory of \cref{action-free massless scalar} has a unitary, causal, Lorentz-invariant, and positive energy UV completion, its Wilson coefficients must satisfy the bound
\begin{equation}\label{upper bound on Wilson coefficients}
    \left|\frac{g_{3,1}}{g_{2,0}}\right| \Lambda^2 \lesssim 5.3,
\end{equation}
where $\Lambda$ is the UV cutoff of the EFT. This constraint comes from the fact that with a unitary, causal, Lorentz-invariant, and positive energy UV completion, ratios of Wilson coefficients like $g_{3,1}\Lambda^2/g_{2,0}$ have to sit in a certain polytope known as the EFT-Hedron \cite{Arkani-Hamed:2020blm,Chiang:2021ziz}. Since that polytope is a bounded region, this allows one to give the upper bound \cref{upper bound on Wilson coefficients}. The same bound has been obtained also outside the context of the EFT-Hedron \cite{Caron-Huot:2020cmc}, using methods such as partial wave unitarity, see, for example \cite{Schwartz:2014sze}, or many other QFT textbooks. Once the Wilson coefficients are fixed, the condition \cref{upper bound on Wilson coefficients} can also be seen as a bound on the UV cutoff. Using \cref{Wilsoncoeffs} we see that in our case it indeed implies that:
\begin{equation} \label{CutoffCond}
    \Lambda^2 \lesssim  5.3 \left|\frac{g_{2,0}}{g_{3,1}}\right| \approx {2.4} \frac{1}{M_P^2 |\lambda_1|}.
\end{equation}
The  UV cutoff of the EFT is thus of the order $1/(\lambda_1 M_P^2)$. If the matter Lagrangian is the Standard Model Lagrangian, the ghost has to be at least heavier than the weak scale so that $\lambda_1 \lesssim 1/(m_W^2 M_P^2)\sim 1/(10^{11} \text{GeV})^4$. As mentioned in the previous paragraphs, for such a small coupling the theory is expected to be almost identical to GR. This confirms that cross-terms of the $R T$ form can only be a possibly consistent modification of GR at very high energy scales.

\subsection{\texorpdfstring{Combining both contributions: $f_0(T)+Rf_1(T)$}{combination}}
\label{zeroth + first order}

One could wonder whether the ghost described in \cref{subsec:matter-curvature-example} can be made heavy by a specific addition of an $f_0(T)$ term. We now investigate this question and consider the action:
\begin{equation}
    S = \int d^4x \sqrt{-g} \left\{ \frac{M_P^2}{2} R f_1(T)+f_0(T) \right\} +\int d^4x \sqrt{-g} \mathcal{L}_m(g_{\mu \nu}, \chi).
\end{equation}
Again for simplicity, we choose $f_1(T)=1+\lambda_1 T$ and  $\mathcal{L}_m$ describing a free massless scalar field. The action in Einstein frame reads:
\begin{align}
     S=\int d^4x \sqrt{-\Tilde{g}} & \left\{ \vphantom{\Big( \frac{X}{X}\Big)} \frac{M_P^2}{2} \Tilde{R}-\frac{3}{4} M_P^2 \lambda_1^2 (1-\lambda_1 X)^2 \p_{\mu}X \Tilde{g}^{\mu \nu} \p_{\nu} X \right. \nonumber\\
     & \left. + \frac{1}{2}(1-\lambda_1 X)X + (1-\lambda_1 X)^2 f_0\Big(\frac{X}{1-\lambda_1 X}\Big)\right\}.
\end{align} 
The presence of the ghost is again related to higher-order derivatives present in the $\nabla_{\mu}X \nabla^{\mu}X$ term. Hence $f_0(T)$ terms cannot eliminate the ghost although they can change its mass.

The ghost scale is related to the ratio of coefficients of the $X^2$ and $\nabla^{\mu}X \nabla_{\mu}X$ term, as it has been explicitly shown in \cref{subsubsec:background-ghost}. Namely, the ghost scale depends directly on the parameter $\gamma$ defined in \cref{alphabetagamma}. In the presence of $f_0(T)$ terms, modifying the coefficient of the $X^2$ term in the Lagrangian, such a parameter can be modified. Indeed, taking for instance $f_0(T)=\gamma_1 T^2/2$, the $X^2$ terms are now proportional to  $(\gamma_1 -\lambda_1)/2$ so that the new parameter reads:
\begin{equation}
    \gamma=\frac{\lambda_1-\gamma_1}{\lambda_1^2M_P^2}=\frac{\lambda_1-\gamma_1}{\lambda_1 }\gamma^{(\gamma_1=0)}. \label{gg1}
 \end{equation}
 
From from the amplitude constraints, one obtains a similar dependence of the cutoff of the EFT with the additional $f_0(T)$ term. Indeed, as stated in \cref{upper bound on Wilson coefficients}, the UV cutoff is given as a ratio of the $X^2$ coefficient and the coefficient $g_{3,1}$, which stays unchanged. We thus obtain the new bound
\begin{equation}
    \Lambda^2 \lesssim 2.4 \frac{|\lambda_1-\gamma_1|}{M_P^2 \lambda_1^2},
\end{equation}
which shows the same dependence in $\gamma_1$ as \cref{gg1}. We see that the only way to increase the ghost scale, or the cutoff of the EFT, compared to the theory of the previous subsection without the $f_0(T)$ term, is thus to have $\gamma_1\gg \lambda_1$. In that case, the effects of the $Rf_1(T)$ term are completely suppressed compared to the $f_0(T)$ one, and the situation effectively reduces to the one described in $\ref{subsec:higher-matter-general}$.

\section{Conclusions and discussion}

\label{sec:conclusions}

This paper is intended to showcase why we deem $f(R,{\rm Matter})$ theories generically irrelevant for cosmological models. In \cref{sec:generalarguments}, we have presented general arguments independent of any specific setup. They should be enough to convince the reader. We then showed how to apply these arguments on models  of the $f(R,{\rm Matter})$. In \cref{sec:specifics}, we gave a concrete example of the appearance of a ghost in a $f(R,T)$ theory. We emphasized that modifications of the type of $f(R,\text{Matter})$ can broadly be categorized into two cases. The first case is a modification of the matter sector. Late-time modifications of such type are strongly constrained by particle physics, as any modification of the matter sector that is significant at energies $\sim H$ must be very strong at the weak scale. The second case is a cross-coupling between curvature and matter. Such coupling generically leads to ghosts, which is excepted to become lighter and lighter as the model deviates from GR.

Before closing this paper, we emphasize that although we focused in this work on $T_{\mu \nu},\mathcal{L}_m$ in terms of fundamental fields (scalar, vector, etc \ldots), our results carry over to the situation where one considers $T_{\mu \nu}$ ($\mathcal{L}_m$) as the energy-momentum tensor of a perfect fluid. The reason for that is simple: the energy-momentum tensor of a perfect fluid should be seen as an effective expression derived from a microscopic description of the matter fields. The canonical example is the energy-momentum tensor of a scalar field, which can be understood as the energy-momentum tensor of a perfect fluid, with the scalar field values determining the pressure and mass density. Nonlinear terms in $f(T)$, when $T$ is a perfect fluid energy-momentum tensor, would also be present in the microscopic theory, giving rise to the problems discussed at length in the previous sections. For example, scattering amplitude constraints are still applicable \cite{deRham:2021fpu}. This immediately generalizes to more complicated situations, e.g. the matter fields in the universe, leading to the same constraints on these models as for their microscopic counterpart. The only way to evade such constraints would be to discover a model in which the problematic terms are absent from the microscopic Lagrangian. Instead, they would be effectively generated when assuming a perfect fluid solution. Such a mechanism is not known to us.

As it is impossible to address every model within the $f(R,\text{Matter})$ class, we encourage advocates of these models to examine whether specific instances somehow circumvent the problems discussed here. This paper aims to clarify why one should not expect such exceptions.

\acknowledgments

The work of SM was supported in part by the World Premier International Research Center Initiative (WPI), MEXT, Japan. SM is grateful for the hospitality of Perimeter Institute, the cosmology group at Simon Fraser University and the Theoretical Physics Institute at University of Alberta, where part of this work was carried out. Research at Perimeter Institute is supported in part by the Government of Canada through the Department of Innovation, Science and Economic Development and by the Province of Ontario through the Ministry of Colleges and Universities.

The work of JS was supported in part by the JASSO scholarship of the state of Japan, by a scholarship of the Friedrich Naumann Foundation for Freedom, and a scholarship of the Max Weber program of the state of Bavaria. JS is grateful for the hospitality of SM and the Yukawa Institute for Theoretical Physics at Kyoto University, where part of this work was carried out.

\bibliography{fRMatter}
\bibliographystyle{JHEP}

\end{document}